\documentclass[preprint,aps,floats,nofootinbib,amssymb,superscriptaddress]{revtex4}
\usepackage[sort&compress]{natbib}
\usepackage{epsf,epsfig}
\usepackage{amsmath}
\usepackage{subfigure}
\usepackage{graphicx}
\usepackage{color}
\usepackage{mathrsfs}

\newcommand{\be}{\begin{equation}}
\newcommand{\ee}{\end{equation}}
\newcommand{\bea}{\begin{eqnarray}}
\newcommand{\eea}{\end{eqnarray}}

\newcommand{\ba}{\begin{array}}
\newcommand{\ea}{\end{array}}
\newcommand{\bi}{\begin{itemize}}
\newcommand{\ei}{\end{itemize}}

\makeatletter

\begin{document}

%-----------------------------------
% Preprint numbers
%-----------------------------------
\preprint{ICCUB-14-001}

%-----------------------------------
% Title
%-----------------------------------
\title{\vspace*{1.in} Photons in a cold axion background and strong magnetic fields: polarimetric consequences}

%-----------------------------------
% Authors
%-----------------------------------
\author{Dom\`enec Espriu}\affiliation{Departament d'Estructura i Constituents de la Mat\`eria,
Institut de Ci\`encies del Cosmos (ICCUB), \\
Universitat de Barcelona, Mart\'i Franqu\`es 1, 08028 Barcelona, Spain}
\author{Albert Renau}\affiliation{Departament d'Estructura i Constituents de la Mat\`eria,
Institut de Ci\`encies del Cosmos (ICCUB), \\
Universitat de Barcelona, Mart\'i Franqu\`es 1, 08028 Barcelona, Spain}

\vspace*{2cm}

\thispagestyle{empty}

%-----------------------------------
% Abstract
%-----------------------------------
\begin{abstract}
In this work we analyze the propagation of photons in an environment where a strong magnetic field
(perpendicular to the photon momenta) coexists with an oscillating cold axion background with the characteristics
expected from dark matter in the galactic halo. Qualitatively, the main effect of the combined
background is to produce a three-way mixing among the two photon polarizations and the
axion. It is interesting to note that in spite of the extremely
weak interaction of photons with the cold axion background, its effects compete with those coming from
the magnetic field in some regions of the parameter space. We determine (with one plausible simplification)
the proper frequencies and eigenvectors as well as
the corresponding photon ellipticity and induced  rotation of the polarization plane that depend both on
the magnetic field and the local density of axions.
We also comment on the possibility that some of the predicted effects could be measured in optical table-top
experiments.
\end{abstract}

\maketitle

\section{Introduction}\label{sec:intro}
Originally introduced to solve the strong $CP$ problem \cite{PQ1,PQ2,PQ3}, axions are an attractive and viable
candidate for dark matter (DM) \cite{sik1,sik2,sik3,raff}. The axion is the Goldstone boson associated with the
spontaneous breaking of the $U(1)_{PQ}$ symmetry \cite{PQ1,PQ2,PQ3}. After the QCD phase transition,
instanton effects induce a potential on the axion field, giving it a mass $m_a$. Astrophysical and
cosmological constraints (see below) force this mass
to be quite small. Yet, the axion provides cold dark matter, as it is not produced thermically.
If the axion background field is initially misaligned (not lying at the bottom of the instanton-induced potential),
at late times it oscillates coherently as
\begin{equation}\label{cos}
 a_b(t)= a_0 \sin m_a t,
\end{equation}
where the amplitude, $a_0$, is related to the initial misalignment angle.
The oscillation of the axion field has an approximately constant (i.e. space-independent) energy
density $\rho=\frac12 a_0^2m_a^2$, which contributes to the total energy of the universe\footnote{This density is not
really constant, as DM tends to concentrate in galactic halos. Nevertheless it is assumed to change
over very large scales, so
for our purposes it suffices to treat it as a constant.}.
This constitutes the cold axion background (CAB for short). There are suggestions that axions could actually
form a Bose-Einstein condensate (BEC) \cite{BEC}.

Axions couple to photons through the term
\begin{equation}\label{lagg}
\mathcal L_{a\gamma\gamma}=g_{a\gamma\gamma}\frac{\alpha}{2\pi}\frac a{f_a}F_{\mu\nu}\tilde F^{\mu\nu},
\end{equation}
where the coefficient $g_{a\gamma\gamma}$ depends on the model\footnote{Sometimes a dimensionful
coupling constant $G_{a\gamma\gamma}\propto\frac1{f_a}$ is used instead.
Our $g_{a\gamma\gamma}$ is dimensionless.}. However, most of them \cite{models1,models2,models3,models4} give $g_{a\gamma\gamma}\simeq 1$. For the present
discussion this is all that matters. The near-universality of the axion-photon coupling makes it the best
candidate to explore axion physics.

This coupling is severely bounded.
The lower limit $f_a > 10^{7}$ GeV coming from astrophysical considerations seems now well established.
If one assumes that axions are the main ingredient of DM, there is also an upper bound:
$f_a < 3\cdot10^{11}\text{ GeV}$. See Ref. \cite{raff} and references
therein for an explanation of the above bounds.
These values of $f_a$  make the axion very weakly coupled and imply a very long lifetime, of the
order of $10^{24}$ years or more; see e.g. Ref. \cite{pdg}.
In the case of Peccei-Quinn axions (i.e. solving the strong $CP$ problem) the approximate
relation $f_am_a\simeq f_\pi m_\pi$ should hold and therefore
cosmology considerations place the axion mass in the range $10^{-1} - 10^{-6}$ eV.
For other axion-like particles, not related to the strong $CP$ problem, there is no such relation
between $m_a$ and $f_a$
and the range of possible values is more open, although they are less motivated from a physical point of view.

Axions could also couple to matter, although in this case the coupling is much more model dependent.
The coupling is so small that their detection is very difficult.
Nevertheless, some of the best bounds on axion masses and couplings come from the study of abnormal
cooling in white dwarfs due to axion emission \cite{wd}.

When dealing with axions and their possible cosmological relevance, there are several separate issues
that have to be adressed. The first one is whether a particle
with the properties of the axion exists or not. This is what experiments such as CAST \cite{cast},
IAXO \cite{iaxo} or ALPS \cite{alps} are addressing directly.
If the axion does exist and its mass happens to be in the relevant range for cosmology
we would have a strong hint that axions may serve as valid DM candidates. Of course for axions to be the
main component of DM
they also have to be present in a sufficient amount.

The mechanism of vacuum misalignment and the
subsequent redshift of momenta suggest that it is natural for axions to remain coherent (or
very approximately so) over relatively long distances, perhaps even forming a BEC as has been suggested.
Thus one should expect not only that the momentum of individual axions satisfies the condition
$k\ll m_a$ as required from cold DM but also all that axions oscillate in phase, rather than incoherently,
at least locally. In addition one needs that the modulus of the axion field is large
enough to account for the DM density.

Finding an axion particle with the appropriate characteristics is not enough
to demonstrate that a CAB exists. Detecting the coherence of the axion background and hence validating
the misalignment proposal is therefore not within reach of any of the above experiments.

The ADMX Phase II experiment \cite{admx} tries to detect axions in the
Galaxy dark matter halo that, under the influence of a strong magnetic field, would convert to photons with
a frequency equal to the axion mass in a resonant cavity. This experiment is sensitive to the local axion density,
the probability of a positive detection being proportional to the latter. In order to get a significant signal the axion
field has to be significantly constant at length scales comparable to the cavity size.
ADMX is therefore sensitive to the CAB.  The experiment claims sensitivity
to axions in the approximate mass range 10$^{-6}$ eV  to 10$^{-5}$ eV and this is also the range of momenta
at which the axion background field can be significantly probed in such an experiment.

Looking for the collective effects on photon propagation resulting from the presence of a CAB is another
possible way of investigating whether a CAB is present at the scales probed by the experiment.
Of course we do not anticipate large or dramatic effects given the presumed smallness of the photon-to-axion
coupling and the low density background that a CAB would provide. However, interferometric and polarimetric
techniques are very powerful and it is interesting to explore
the order of magnitude of the different effects in this type of experiments.
Potentially, photons can also probe the CAB structure
in different ranges of momenta. In addition, precise photon measurements could in principle
check the coherence of the oscillations over a variety of distances.
Discussing in detail the effects of a CAB on photons is the purpose
of the present paper.

Several studies on the influence of axions on photon propagation at cosmological scales exist\cite{cosmoaxions1,cosmoaxions2,cosmoaxions3,cosmoaxions4}.
The consequences are only visible for extremely low mass axions, such as the ones hypothetically
produced in string theory scenarios\cite{axiverse}.
We do not consider very light axions here in detail as their masses do not fall into the favoured range but
exploring such small masses might be of interest too.

It is worth noting that a CAB introduces via its time dependence some amount of  Lorentz-invariance violation
in photon physics; the term \eqref{lagg} does actually modify
the photon dispersion relation and it has somewhat exotic consequences. For instance, in Ref. \cite{shield} we
showed how this modification of the dispersion relation allows the emission of a photon by a cosmic ray,
a process forbidden due to conservation of energy and momentum in a Lorentz-invariant theory.
In Ref. \cite{cr} we computed the amount of energy radiated by this process and found it to be non-negligible,
although the normal synchrotron radiation background makes its detection very challenging.

In Ref. \cite{last} we found that some photon wave numbers are actually forbidden in a CAB, as a consequence
of its time dependence. This striking result is an unavoidable and direct consequence of the periodicity
of the CAB oscillations. In the subsequent we study the actual width of these gaps that, not
surprisingly, turns out to be extremely narrow for axions of cosmological relevance.

The consequences
of the mere existence of axions as propagating degrees of freedom on
photon propagation have been studied for a long time and are well understood.
It is well known that photons polarized in
a direction perpendicular to the magnetic field are not affected by the existence of axions \cite{raffelt1,raffelt2} but
photons polarized in the parallel direction mix with them. As a consequence there is a small rotation in the
polarization plane due to photon-axion mixing as well as a change in the ellipticity \cite{mpz}.
In Ref. \cite{last} we showed that a similar effect exists even without
a magnetic field when a CAB is present except that now it involves
the two photon helicities.

Throughout this work we will see that the effects of the CAB on the propagation of photons are extremely small,
so it is quite pertinent to question whether these effects could be experimentally measured. The answer is surely
negative with present day experimental capabilities but some effects are not ridiculously small either to be
discarded from the outset: the effects of a coherent CAB are in some cases quite comparable
to, or even larger than, the influence of axions as mere propagating degrees of freedom, which
have been profusely studied before. They might even be comparable
to non-linear QED effects, which have also been actively sought for experimentally. Therefore we think
it is legitimate to present this study in view of the physical relevance of the presumed existence of a CAB
as a dark matter candidate.

This work is a continuation of Ref. \cite{last} and some overlap is unavoidable to be reasonably self-contained.
In section \ref{sec:eom} we review the problem and derive the equations
of motion for the axion and photon in the presence of both backgrounds, both for linear and circular polarization
 bases for the photon. We also review there the range of relevant values for the intervening parameters.
In section \ref{sec:gaps} we discuss the results for the case of no magnetic field, when there is no photon-axion
conversion but the CAB still mixes the two photon helicities. In Ref. \cite{last} we found that some
gaps in the photon momenta were present due to the time periodicity of the CAB.
We complement this discussion now by deriving the precise location and width of these momentum
gaps. In section \ref{sec:magnetic} we study the consequences that the combined background
has on photon wave-numbers and polarizations. In section \ref{sec:rotation} we
explore the consequences of the change in the plane of polarization of the photons in the presence of the CAB, making
use of the photon propagator derived in a combined CAB and constant magnetic field that was derived in Ref. \cite{last}.
We also correct some approximations that were made in Ref. \cite{last} and that turn out not to be correct for the
relevant range of masses, magnetic fields and CAB densities. We have also made an effort in presenting the new
the results in a standard notation, more suitable for optical experiments.
The technical details are given in a detailed appendix. A short summary of some of our results was already presented in
Ref. \cite{mainz}.

\section{Equations of motion of the axion-photon system}\label{sec:eom}
The Lagrangian density describing axions and photons consists of the usual kinetic terms plus
the interaction term \eqref{lagg}
\begin{equation}\label{fulllagrangian}
\mathcal{L}=\frac12\partial_\mu a\partial^\mu a-\frac12m_a^2a^2-\frac14F_{\mu\nu}F^{\mu\nu}+
\frac g4aF_{\mu\nu}\tilde F^{\mu\nu},
\end{equation}
where we have rewritten the axion-photon coupling
%\footnote{In the literature one often finds this dimensionful coupling written as $G_{A\gamma\gamma}$.}
as $g=g_{a\gamma\gamma}\frac{2\alpha}{\pi f_a}$.
We are not considering the non-linear effects due to the Euler-Heisenberg Lagrangian \cite{EH1,EH2}
that actually can provide some modifications in the polarization plane.
Later we shall discuss their relevance.

We decompose the fields as a classical piece describing the backgrounds
(external magnetic field $\vec B$  and a CAB as given in \eqref{cos} plus quantum fluctuations describing
the photon and the axion particles, e.g. $a\to a_b + a$.
For the (quantum) photon field, we work in the Lorenz gauge,
$\partial_\mu A^\mu=0$, and use the remaining gauge freedom to set $A^0=0$. The equations of motion are
\begin{equation}\label{EL}
\begin{array}{l}
(\partial_\mu\partial^\mu+m_a^2)a+gB^i\partial_tA_i=0,\\
\partial_\mu\partial^\mu A^i+gB^i\partial_ta+\eta\epsilon^{ijk}\partial_jA_k=0,
\end{array}
\end{equation}
where $\eta=g\partial_t a_b$. We neglect the space derivatives of $a_b$ thereby assuming homogeneity
of the axion background, at least at the scale of the photon momentum and translational invariance.
Since $\eta$ is time-dependent, we make a Fourier transform with respect to
the spatial coordinates only,
\begin{equation}
\phi(t,\vec x)=\int\frac{d^3k}{(2\pi)^3}e^{i\vec k\cdot\vec x}\hat\phi(t,\vec k),
\end{equation}
and get the equations
\begin{equation}\label{TF}
\begin{array}{l}
(\partial_t^2+\vec k^2+m_a^2)\hat a+gB^i\partial_t \hat A_i=0,\\
(\partial_t^2+\vec k^2)\hat A^i+gB^i\partial_t\hat a+i\eta\epsilon^{ijk}k_j\hat A_k=0.
\end{array}
\end{equation}
As can be seen, the presence of a magnetic field mixes the axion with the photon.
To proceed further, we write the photon field as
\begin{equation}
 \hat A_\mu(t,\vec k)=\sum_\lambda f_\lambda(t)\varepsilon_\mu(\vec k,\lambda),
\end{equation}
where $\varepsilon_\mu$ are the polarization vectors and $f_\lambda(t)$ are the functions we will have to solve for.
If we choose a linear polarization basis for the photon, the equations are, in matrix form,
\begin{equation}\label{linear}
\left(
\begin{array}{ccc}
 \partial_t^2+k^2+m_a^2 &   -ib\partial_t    & 0        \\
 -ib\partial_t          & \partial_t^2+k^2 & -\eta(t)k                    \\
 0      & -\eta(t)k                & \partial_t^2+k^2
\end{array}
\right)
\left(
\begin{array}{c}
 \hat a \\ i f_\parallel \\ f_\perp
\end{array}
\right)
=
\left(
\begin{array}{c}
 0 \\ 0 \\ 0
\end{array}
\right),
\end{equation}
where $k=|\vec k|$ and $b=g|\vec B^\perp|$, where $\vec B^\perp$ is the component of the magnetic field
perpendicular to the momentum (the parallel
component does not affect propagation at all if the Euler-Heisenberg piece is neglected).
The subscripts $\parallel$ and $\perp$ refer to parallel or perpendicular to this $ \vec B^\perp$.

In a circular polarization basis, defining
\begin{equation}\label{imaginary}
 f_\pm=\frac{f_\parallel\pm if_\perp}{\sqrt2},
\end{equation}
the equations take the form
\begin{equation}\label{circulareq}
\left(
\begin{array}{ccc}
 \partial_t^2+k^2+m_a^2 &   i\frac b{\sqrt2}\partial_t    & i\frac b{\sqrt2}\partial_t        \\
 i\frac b{\sqrt2}\partial_t   & \partial_t^2+k^2+\eta(t)k & 0                    \\
 i\frac b{\sqrt2}\partial_t      & 0                & \partial_t^2+k^2-\eta(t)k
\end{array}
\right)
\left(
\begin{array}{c}
 i\hat a \\  f_+ \\ f_-
\end{array}
\right)
=
\left(
\begin{array}{c}
 0 \\ 0 \\ 0
\end{array}
\right).
\end{equation}
As we see from the previous expressions, the presence of a CAB changes in a substantial way the mixing
of photons and axions. Now all three degrees of freedom are involved.

A difference in the approach between this work and Ref. \cite{raff&sto} is worth noting.
In going from \eqref{EL} to \eqref{TF} we have performed a
Fourier transform in space, but not in time, because the magnetic field is homogeneous
but $\eta(t)$ is time-dependent. Equation (4) in Ref. \cite{raff&sto}, however,
uses a transform in time rather than in space because the CAB is not considered.

\begin{figure}[ht]
 \center
 \includegraphics[scale=0.8]{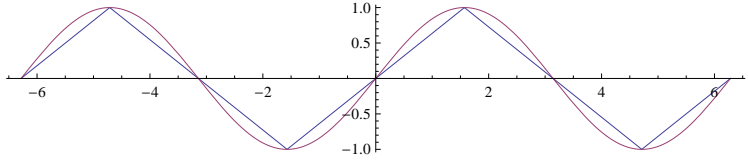}
 \caption{$a_b(m_at)/a_0$ and its approximating function.}\label{fig:tri}
 \end{figure}

There are several ways to deal with the periodic CAB. One possibility is to try to treat it exactly. Unfortunately
this unavoidably leads to the appeareance of Mathieu functions due to the sinusoidal variation
of the background and the analysis becomes extremely involved.
On the other hand, the substantial ingredient in the problem is the existence of periodicity itself
and the fine details are not so relevant\footnote{Recall that the generic appeareance of bands in the energy levels
of a solid relies on the
periodicity of the potential and not on its precise details.}. Therefore to
keep the discussion manageable, we approximate the sinusoidal variation of the axion background $a_b(t)$ in
\eqref{cos} by a piecewise linear function, see figure \ref{fig:tri}.
Since $\eta(t)$ is proportional to the time derivative of $a_b(t)$, in this approximation it is a square-wave function,
alternating between intervals
where $\eta=\eta_0$ and $\eta=-\eta_0$ with a period $2T=2\pi/m_a$.
Here, $\eta_0=\frac2\pi ga_0m_a=g_{a\gamma\gamma}\frac{4\alpha}{\pi^2}\frac{a_0m_a}{f_a}$.

A brief numerical discussion of the parameters involved in the problem and their relative
importance is now in order.
The bound on $f_a$ implies one on $g$. Taking $g_{a\gamma\gamma}$ of $\mathcal{O}(1)$ the
range $f_a=10^7 - 10^{11} \text{ GeV}$  translates to $g=10^{-18} - 10^{-22}\text{ eV}^{-1}$.
Assuming a halo DM density of $\rho=10^{-4}\text{ eV}^4$ \cite{halo} this means that $\eta_0=10^{-20} - 10^{-24} \text{ eV}$.
When working with natural units and magnetic fields it is useful to know that $1\text{ T}\approx195\text{ eV}^2$.
To have a reference value, a magnetic field of $10\text{ T}$ implies the
range $b=10^{-15} - 10^{-19}\text{ eV}$, for $f_a=10^7 - 10^{11} \text{ GeV}$.

Finally, let us now comment on the relevance of the contribution of the Euler-Heisenberg pieces compared
to the ones retained
in the description provided by \eqref{fulllagrangian}. As it is known (see e.g. Ref. \cite{raffelt1,raffelt2}) an external
magnetic field perpendicular to the photon motion contributes, via the Euler-Heisenberg terms, to the
mixing matrices,  affecting the (2,2) and (3,3) entries of \eqref{linear} and \eqref{circulareq}.
They modify the $k^2$ terms with corrections of order $10^{-2}\times \alpha^2 \times (B^2/m_e^4)$, where
$m_e$ is the electron mass, leading to birefringence and therefore to ellipticity. For magnetic fields of $\sim 10$ T this
gives a contribution of order $10^{-21}$ that
may be comparable to axion-induced effects for large magnetic
fields, particularly if $f_a$ is very large, or to the effects from the CAB (which for $k\sim 1$ eV are
in the range $10^{-20} - 10^{-24}$). Since there is no new physics involved
in the contribution from the Euler-Heisenberg Lagrangian, in order to facilitate the analysis we will not consider it here.
In any case given the smallness of the Euler-Heisenberg and the axion effects, they can safely be
assumed to be additive \cite{raffelt1,raffelt2}. The relevant modifications due to
the Euler-Heisenberg term can be found in Refs. \cite{raffelt1,raffelt2} and \cite{pvlas2}.

Of course, the effects of the Euler-Heisenberg Lagrangian are absent or negligible if there is no magnetic field
or if it is relatively weak,
and we will see that for a range of parameters the effect of a CAB  might be comparable to the former.

\section{No magnetic field: forbidden wavelengths}\label{sec:gaps}
If there is no magnetic field ($b=0$, $\eta_0\neq0$) the axion and the photon are no longer mixed.
We explored this situation in Ref. \cite{last} and we will summarize the main results here and
complete the discussion.

\begin{figure}[tbh]
\centering
\subfigure[(a)]   {\includegraphics[clip,width=0.35\textwidth]{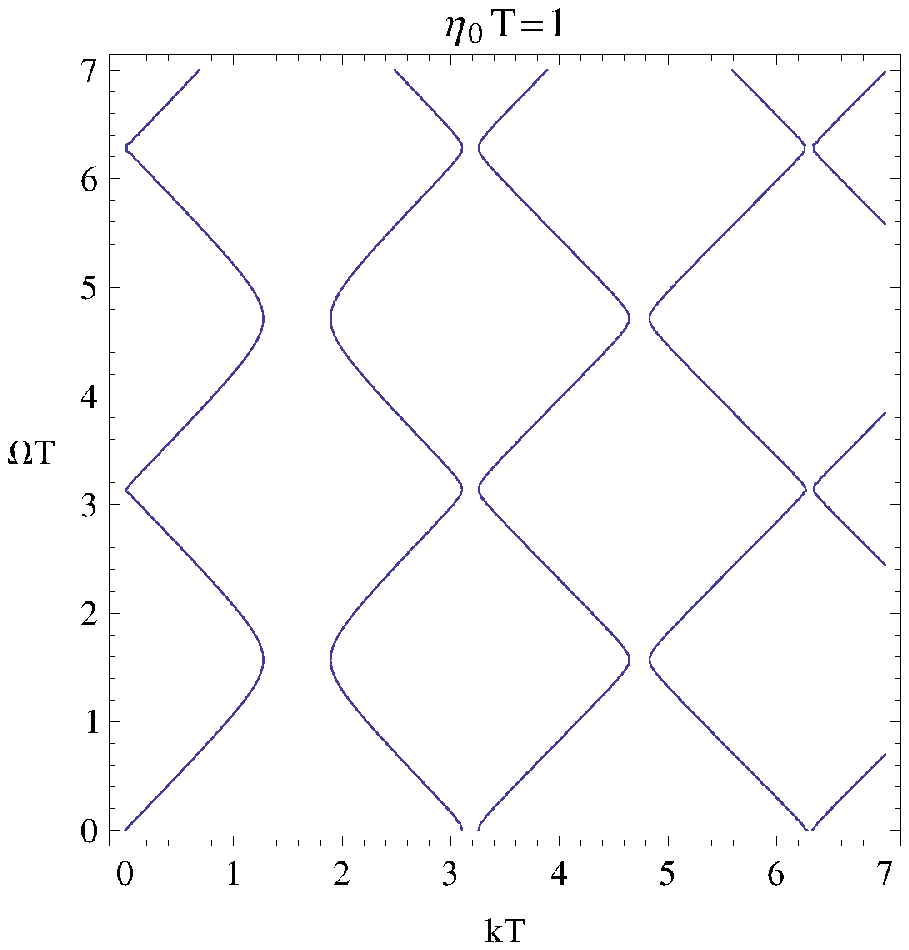}} \hspace{1cm}
\subfigure[(b)]{\includegraphics[clip,width=0.35\textwidth]{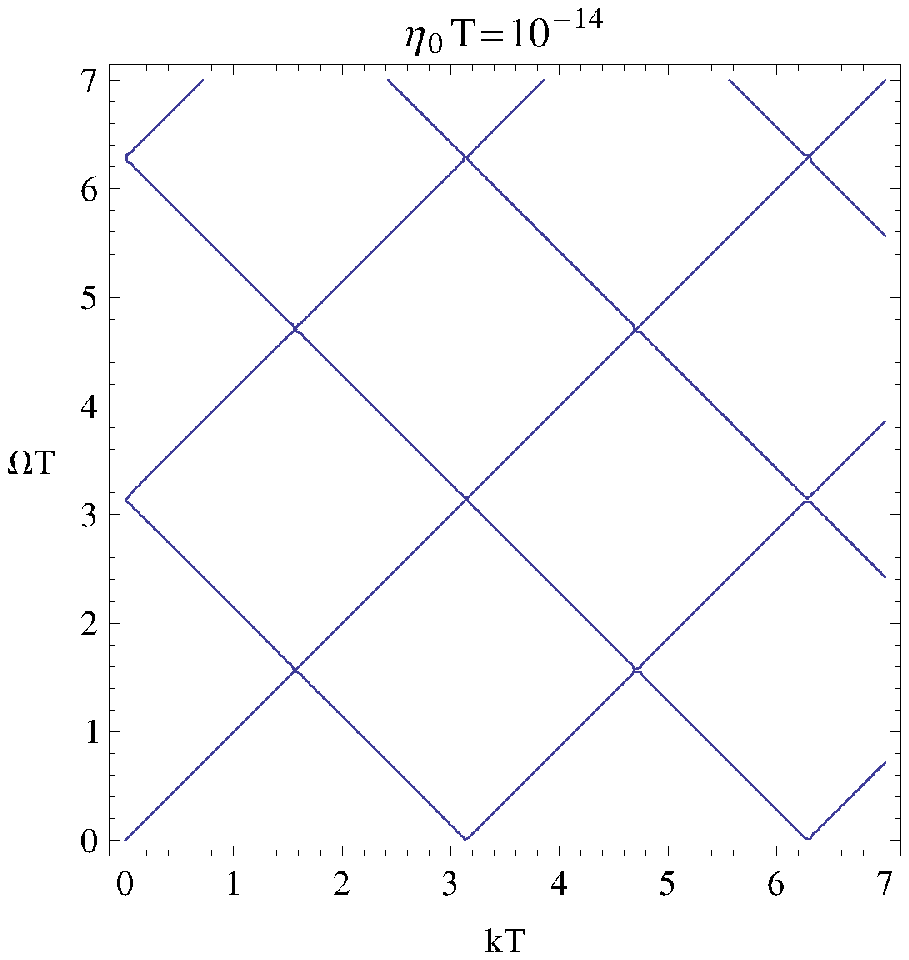}} \\
\caption{Plot of the solutions to the gap equation. In the left figure the value for the ratio $\eta_0/m_a$
is unreasonably large
and it is presented here only to make the gaps in the photon momentum clearly visible.}
\label{fig:gaps}
\end{figure}

Because $\eta(t)$ does mix the two linear polarizations, in this case it is useful to choose the circular polarization
basis, which diagonalizes the system.
The solution is, for a given interval, $f_\pm(t)=e^{i\omega_\pm t}$, $\omega_\pm=\sqrt{k^2\pm\eta_0 k}$. Of course, when
$\eta(t)$ changes sign the solutions are interchanged as well. The way to solve this is to
write $f_\pm(t)=e^{i\Omega t}g_\pm(t)$ and demand that $g_\pm(t)$ have the same periodicity as $\eta(t)$. After
elementary quantum mechanical considerations, periodicity of the modulus of the wave-function
imposes the condition
\begin{equation}\label{eq:gaps}
 \cos(2\Omega T)=\cos(\omega_+ T)\cos(\omega_-T)-\frac{\omega_+^2+\omega_-^2}{2\omega_+\omega_-}\sin(\omega_+ T)\sin(\omega_-T).
\end{equation}
This condition implies the existence of momentum gaps: some values of $k$ admit no solution for $\Omega$, much like
some energy bands are forbidden in a semiconductor. Here, however, the roles of momentum and energy are
exchanged, since the periodicity is in time,
rather than in space.
The solutions are shown in an $\Omega(k)$ plot in figure \ref{fig:gaps} for two values of the ratio $\eta_0/m_a$.
One of the ratios shown is unreasonably large, in order to
show clearly the existence of the gaps.

Let us now discuss the width of these gaps, an issue that was not
studied in Ref. \cite{last} in detail.
The first order in $\eta_0$ drops from \eqref{eq:gaps} but to second order it reads
\begin{equation}
 \cos(2\Omega T)=\cos(2kT)+\frac{\eta_0^2}{4k^2}\left[-1+\cos(2kT)+kT\sin(2kT)\right]
\end{equation}
(recall that $T=\pi/m_a$). There is no solution when the r.h.s. of this expression becomes larger
than one. The gaps are approximately located at
\be
k_n=\frac{nm_a}2,~n\in\mathbb{N}
\ee
and their width is
\begin{equation}
 \Delta k\sim\left\{\begin{array}{cc}
\displaystyle\frac{\eta_0}{n\pi} & \text{for $n$ odd} \\
\\
\displaystyle\frac{\eta_0^2}{2nm_a} & \text{for $n$ even}
                    \end{array}
                    \right..
\end{equation}
These results agree well with the exact results as can be easily seen in the left side of figure \ref{fig:gaps}.
Unfortunately
we are not aware of any way of detecting such a tiny forbidden band for the range of values of $\eta_0$ previously
quoted ($10^{-20}$ eV or less) that correspond to the allowed values of $f_a$.

It may be interesting to think what would happen if one attempts to produce a `forbidden' photon, i.e. one
whose momentum falls in one of the forbidden bands. A photon with
such a wave number is `off-shell' and as such it will always decay. For instance, it could decay into three
other photons with appropiately lower energies. However, because the off-shellness is so small (typically
$10^{-20}$ eV or less) it could live for a long time as a metastable state, travelling distances
commensurable with the solar system. For more technical details see e.g. Ref. \cite{sasha}.

We realize that the small bandwidth of the forbidden momentum bands make them unobservable
in practice. However
their mere existence is of theoretical interest. Conclusions might be different for other axion-like backgrounds.

\section{Proper modes in a magnetic field and axion background}\label{sec:magnetic}
In the presence of a magnetic field, but no CAB ($b\neq0$, $\eta_0=0$) there is no longer a
time dependence in the coefficients of the equations, so
we can Fourier transform with respect to time as well. We find the following dispersion relations:
\begin{eqnarray}
\omega_a^2&=&k^2+\frac{m_a^2+b^2}{2}+\frac12\sqrt{(m_a^2+b^2)^2+4b^2k^2}\approx (k^2 + m_a^2)
\left(1+\frac{b^2}{m_a^2}\right)\cr
\omega_1^2&=&k^2+\frac{m_a^2+b^2}{2}-\frac12\sqrt{(m_a^2+b^2)^2+4b^2k^2}\approx k^2 \left(1-\frac{b^2}{m_a^2}\right)\cr
\omega_2^2&=&k^2,
\end{eqnarray}
where the $\approx$ symbol indicates the limit $\frac{bk}{m_a^2}\ll1$. These results are well known \cite{mpz}.
We have identified as corresponding to `photons' the two modes that if $b=0$ reduce to the two usual polarization
modes. The third frequency corresponds predominantly
to the axion (or axion-like particle), but of course it has also a small photon component as
the $\parallel$ polarized photon mixes with the axion.

If laser light of frequency $\omega$ is injected into a cavity, the different components will develop different
wave-numbers resulting in the appeareance of changes in the plane of polarization (ellipticity
and rotation) unless the photon polarization is initially exactly parallel or exactly perpendicular to the
magnetic field. We will review these effects later.
From the above expressions it would appear that the relevant figure of merit to observe
distortions with respect the unperturbed photon propagation is the ratio $\frac{b^2}{m_a^2}$ and
this is indeed true at large times or distances (actually for $x\gg \frac{\omega}{m_a^2}$). This number is of
course very small, typically $10^{-28}$ for the largest conceiveable magnetic fields (note that this
ratio is actually independent of $f_a$ and $m_a$ provided that we are considering Peccei-Quinn axions.)

Laser interferometry is extremely precise and Michelson-Morley type experiments
are capable of achieving a relative error as small as $10^{-17}$ using  heterodyne interferometry techniques\cite{LI1,LI2}
and the PVLAS collaboration  claims that a sensitivity of order $10^{-20}$ in the difference of
refraction indices is ultimately achievable \cite{pvlas} (see also Ref. \cite{tamyang}).
In spite of this the above figure seems way too small to be detectable.

Let us now explore the situation where both the CAB and the magnetic field are present. We choose to work with the
linear polarization basis. Again, in each time interval  we can define $(a,if_\parallel,f_\perp)=e^{i\omega t}(x,iX_\parallel,X_\perp)$.
Then the equations in matrix form are
\begin{equation}
\left(
\begin{array}{ccc}
 -\omega^2+k^2+m_a^2 & \omega b         & 0         \\
 \omega b          & -\omega^2+k^2 & -\eta_0 k                    \\
 0      & -\eta_0 k                & -\omega^2+k^2
\end{array}
\right)
\left(
\begin{array}{c}
x\\ iX_\parallel \\ X_\perp
\end{array}\right)
=
\left(
\begin{array}{c}
 0 \\ 0 \\ 0
\end{array}
\right),
\end{equation}
and involve a full three-way mixing as previously mentioned.
The proper frequencies of the system turn out to be
\begin{eqnarray}
\omega_a^2&=& k^2+\frac{m_a^2+b^2}3+2\sqrt{Q}\cos\phi,\cr
\omega_1^2&=& k^2+\frac{m_a^2+b^2}3-\sqrt{Q}\left(\cos\phi+\sqrt3\sin\phi\right),\cr
\omega_2^2&=& k^2+\frac{m_a^2+b^2}3-\sqrt{Q}\left(\cos\phi-\sqrt3\sin\phi\right),
\end{eqnarray}
where
\begin{eqnarray}
 Q&=&\left(\frac{m_a^2+b^2}3\right)^2+\frac13k^2(b^2+\eta_0^2),\cr
 \phi&=&\frac13\arctan\frac{\sqrt{Q^3-R^2}}R,\cr
 R&=&\frac1{54}(m_a^2+b^2)\left[2m^4+b^2(9k^2+4m_a^2+2b^2)\right]-\frac16\eta_0^2k^2(2m_a^2-b^2).
\end{eqnarray}
It can be observed that they depend only on even powers of $\eta_0$, so they are not altered when $\eta(t)$ changes sign.
According to the discussion at the end of section \ref{sec:eom} the limit $\eta_0\ll b\ll\{m_a,k\}$ is quite reasonable.
The approximate expressions for the proper frequencies in this limit are\footnote{Extreme care has
to be exercised when using approximate formulae based on series expansions in $b$ or $\eta_0$ because there
is a competition among dimensionful quantities, several of which take rather small values.}

\begin{eqnarray}
\omega_a^2&\approx& (k^2+m_a^2)\left(1+\frac{b^2}{m_a^2}\right),\cr
\omega_1^2&\approx& k^2-k\sqrt{\eta_0^2+\left(\frac{b^2k}{2m_a^2}\right)^2%\left[1+\frac{2b^2}{m^2}\left(1+\frac{2k^2}{m^2}\right)\right]
}-\frac{b^2k^2}{2m_a^2},\cr
\omega_2^2&\approx& k^2+k\sqrt{\eta_0^2+\left(\frac{b^2k}{2m_a^2}\right)^2%\left[1+\frac{2b^2}{m^2}\left(1+\frac{2k^2}{m^2}\right)\right]
}-\frac{b^2k^2}{2m_a^2}.
\end{eqnarray}

Corresponding to each frequency, the eigenvectors that solve the system are
\begin{equation}\label{autovec}
\omega_a:~
\left(
\begin{array}{c}
1\\
\displaystyle \frac{b\sqrt{k^2+m_a^2}}{m_a^2}\\
\displaystyle -\frac{\eta_0bk\sqrt{k^2+m_a^2}}{m_a^4}
\end{array}\right)
,\quad
\omega_1:~
\left(
\begin{array}{c}
\displaystyle-\frac{bk}{m_a^2}\\
1\\
\varepsilon
\end{array}\right),\quad
\omega_2:~
\left(
\begin{array}{c}
\displaystyle\frac{bk}{m_a^2}\varepsilon\\
-\varepsilon\\
1
\end{array}\right),
\end{equation}
where
\be\label{ellipt}
\varepsilon=\frac{\eta_0}{\sqrt{\eta_0^2+\left(\frac{b^2k}{2m_a^2}\right)^2}+\frac{b^2k}{2m_a^2}}.
\ee

Note that the above eigenvectors are written in the basis described in \eqref{imaginary} that includes an imaginary unit for the
parallel component. Therefore the eigenvectors for $\omega_{1,2}$ correspond to photon states elliptically polarized with
ellipticity
\footnote{Ellipticity is the ratio of the minor to major axes of the ellipse.} $|\varepsilon|$. In addition, unless exactly aligned
to the magnetic field there will be a change in the angle of polarization. We will return to this in section \ref{sec:rotation}.

We also note that the above value for $\varepsilon$ corresponds to the ellipticity of the eigenmodes. In section \ref{sec:rotation}
we will discuss the
evolution of the ellipticity of photon state that is initially linearly polarized.

Let us now try to get some intuition on the relevance of the different magnitudes entering in the expressions.
There are two different limits we can study, depending of which term in the square root in \eqref{ellipt} dominates.
If $\displaystyle\frac{|\eta_0|}{k}\ll \frac{b^2}{2m_a^2}$ we have $\displaystyle\varepsilon\approx\frac{\eta_0m_a^2}{b^2k}$.
The ellipticity of the eigenmodes is small, so the proper modes are almost linearly polarized photons.
In the case $\displaystyle\frac{|\eta_0|}{k}\gg \frac{b^2}{2m_a^2}$ we have
$\displaystyle\varepsilon\approx\text{sign}(\eta_0)\left(1-\frac{b^2k}{2|\eta_0|m_a^2}\right)$.
Now the ellipticity of the eigenmodes is close to $1$ so the proper modes are almost circularly polarized. We see that while the
proper frequencies depend only on the square of $\eta_0$ (and therefore do not change as we go from one
time interval to the next) the eigenvectors do change.

The discussion on the size of the different parameters done in section \ref{sec:eom} and also in this section indicates that
the effect from the cold axion background is actually the dominant one for Peccei-Quinn axions, well and above the
effects due to the presence of the magnetic field. Unfortunately both are minute.
In the limit where the magnetic field can be neglected, the photon proper frequencies are
\be
\omega_\pm^2 = k^2 \pm k \eta_0
\ee

Axion-like particles are not constrained by the PCAC relation $f_a m_a \simeq {\rm constant}$
required of Peccei-Quinn axions and using (somewhat arbitrarily) the
largest value of $b$ discussed and the smallest mass for $m_a$ we get
a value for $b^2/m_a^2$ in the region $\sim10^{-18}$, to be compared with the largest
acceptable value for $\eta_0$ that gives  $\eta_0/k \sim 10^{-20}$ if $k\sim 1$ eV.
Sensitivity to the magnetic field could be enhanced by being able to
reproduce the experiment with even larger magnetic fields.\footnote{Non-destructive magnetic fields close to
$100\text{ T}$ have been achieved. This would enhance the sensitivity by a factor 100.}

\section{Change of the polarization in an axion background}\label{sec:rotation}

For our purposes it will be useful
to consider the electric field correlator, easily derived from the resummed  photon propagator
derived in Ref. \cite{last}. Using that $\vec k\cdot \vec B=0$, we get in momentum space
\be
D^E_{ij}(\omega,k)=-\frac{ig_{ij}\omega^2}{\omega^2-k^2}-\frac{i\omega^4 b_i b_j}
{(\omega^2-k^2)[(\omega^2-k^2)(\omega^2-k^2-m_a^2)-\omega^2 b^2]}.
\ee
Notice the rather involved structure of the dispersion relation implied in the second term, which
is only present when $b\neq 0$, while the first piece corresponds to the unperturbed propagator.
For a given value of the wave-number $k$ the zeros
of the denominator are actually the proper frequencies $\omega_a,\omega_1$ and $\omega_2$.
We consider the propagation of plane waves moving in the $\hat x$ direction.
The Fourier transform with respect to the spatial component will describe the space evolution of
the electric field. We decompose
\be
\frac1{(\omega^2-k^2)[(\omega^2-k^2)(\omega^2-k^2-m_a^2)-\omega^2b^2]}=
\frac{A}{k^2-\omega^2}+\frac{B}{k^2-F^2}+\frac{C}{k^2-G^2},
\ee
where $\omega$, $F$ and $G$ are the roots of the denominator
\begin{eqnarray}
F^2&=&\omega^2-\frac{m_a^2}2+\frac12\sqrt{m_a^4+4\omega^2b^2}\approx\left(1+\frac{b^2}{m_a^2}\right)\omega^2,\cr
G^2&=&\omega^2-\frac{m_a^2}2-\frac12\sqrt{m_a^4+4\omega^2b^2}\approx\left(1-\frac{b^2}{m_a^2}\right)\omega^2-m_a^2,
\end{eqnarray}
and
\begin{eqnarray}
A&=&-\frac1{\omega^2-F^2}\frac1{\omega^2-G^2}=\frac1{\omega^2 b^2},\cr
B&=&-\frac1{F^2-\omega^2}\frac1{F^2-G^2}\approx-\frac1{\omega^2 b^2}\left(1-\frac{\omega^2b^2}{m_a^4}\right),\cr
C&=&-\frac1{G^2-\omega^2}\frac1{G^2-F^2}\approx-\frac1{m_a^4}.
\end{eqnarray}
The last contribution to $B$ in the previous formula was incorrectly neglected in our previous publication
\cite{last}.
The space Fourier transform of the electric field propagator is
\bea
D_{ij}^E(\omega,x)&=&-g_{ij}\frac{\omega}{2}e^{i\omega x}+
\frac{\omega^4}{2}b_ib_j\left(\frac A\omega e^{i\omega x}+ \frac{B}{F}e^{iFx}+\frac{C}{G}e^{iGx}\right)\cr
&=&\frac\omega2e^{i\omega x}\left[-g_{ij}+\omega^3b_ib_j\left(\frac A\omega + \frac{B}{F}e^{i(F-\omega)x}+\frac{C}{G}e^{i(G-\omega)x}\right)\right],
\eea
where $x$ is the travelled distance. After factoring out the exponential $e^{i\omega x}$ we consider the
relative magnitude of the differential
frequencies  $F-\omega$ and $G-\omega$. The latter
is much larger and for $m_a^2 x/2\omega \gg 1$ the corresponding exponential could be dropped. This approximation was made in
Ref. \cite{last} and for the range of axion masses envisaged here and $\omega\sim 1$ eV is valid for all astrophysical
and most terrestrial experiments. As for the exponential containing $F-\omega$, we can safely expand it for table-top experiments
and retain only the first non-trivial term.
In this case, the leading terms in the propagator are
\be
D_{ij}^E(\omega,x)\approx\frac\omega2e^{i\omega x}\left[-g_{ij}+\hat b_i\hat b_j\left(\frac{\omega^2b^2}{m_a^4}-
i\frac{\omega b^2x}{2m_a^2}\right)\right],
\ee
where $\hat b$ is a unitary vector in the direction of the magnetic field.
For very light axion masses, neglecting the $e^{i(G-\omega)x}$ exponential cannot be justified for table top experiments.
Then one should use
a slightly more complicated propagator, namely
\be
D_{ij}^E(\omega,x)\approx\frac\omega2e^{i\omega x}\left\{-g_{ij}+\hat b_i\hat b_j
\frac{\omega^2b^2}{m^4_a}\left[ 1-\cos\frac{m_a^2x}{2\omega} +i\left(\sin\frac{m_a^2x}{2\omega}- \frac{m_a^2x}{2\omega}\right)\right]\right\}.
\ee
These expressions agree in the appropriate limits with the ones in Ref. \cite{mpz}.

When a CAB is considered the electric field propagator changes to
\bea
D^E_{ij}(\omega,k)&=&-i\omega^2\left(\frac{P_{+ij}}{\omega^2-k^2-\eta_0k}+\frac{P_{-ij}}{\omega^2-k^2+\eta_0k}\right)\cr
&&-i\omega^4\frac{ b_i b_j}
{(\omega^2-k^2)[(\omega^2-k^2)(\omega^2-k^2-m_a^2)-\omega^2 b^2]}.
\eea
The $P+$ and $P_-$ are projectors defined in Ref. \cite{sasha}.
This expression differs from the one presented in formula (75) of Ref. \cite{last} in that (a) only the leading
contribution to the term proportional to the magnetic field is retained and (b) the piece independent of the magnetic
field contains (unlike in Ref. \cite{last}) the modifications from the CAB. See the appendix for a complete discussion.
The external magnetic field can be set to zero in the previous expressions, if desired.

By projecting on suitable directions and taking the modulus square of the resulting quantity,
the following expression for the angle of maximal likelihood (namely, the one where it is more
probable to find the direction of the rotated electric field) as a function of the distance $x$ can be found
\be\label{likelyrotat}
\alpha(x)=\beta-\frac{\eta_0x}{2}-\frac\epsilon2\sin2\beta,
\ee
where $\beta$ is the initial angle that the oscillation plane of the electric field
forms with the background magnetic field and
\be
\epsilon\approx-\frac{\omega^2b^2}{m_a^4}\left(1-\cos\frac{m_a^2x}{2\omega}\right).
\ee
From the results in the appendix, the ellipticity turns out to be
\be
e=\frac12\left|\varphi\sin2\beta\right|,\quad \varphi\approx\frac{\omega^2b^2}{m_a^4}\left(\frac{m_a^2x}{2\omega}-\sin\frac{m_a^2x}{2\omega}\right).
\ee
For small distances, $\frac{m_a^2x}{2\omega}\ll1$ we can expand the trigonometric functions to get
\be
\epsilon\approx-\frac{b^2x^2}{8},\quad\varphi\approx\frac{m^2b^2x^3}{48\omega}.
\ee
If this limit is not valid, we have instead
\be
\epsilon\approx-\frac{\omega^2b^2}{m^4},\quad\varphi\approx\frac{\omega b^2x}{2m^2}.
\ee
It can be noted that the effect of the magnetic field always comes with the factor $\sin2\beta$, which means that
it disappears if the electric field is initially parallel
($\beta=0$) or perpendicular ($\beta=\pi/2$) to the external magnetic field.

The results of Ref. \cite{mpz}, which we reproduce in the case where $\eta_0=0$,  are known to be in agreement
with later studies such as Ref. \cite{raff&sto}, which has somehow become a standard reference in the
field. However, their
approach is not adequate to deal with time dependent backgrounds and therefore it is
not easy to reinterpret the results derived in the present work when a non-vanishing CAB is present in the
language of Ref. \cite{raff&sto}.

\section{Measuring the CAB in polarimetric experiments}\label{sec:polarization}

If $\eta_0\neq 0$ a rotation is present even in the absence of a magnetic field. This is a characteristic
footprint of the CAB. This `anomalous' rotation attempts to bring the initial polarization plane to agree
with  one of the two elliptic eigenmodes. In the case where the effect of $\eta_0$ dominates, the eigenmodes are
almost circularly, rather than linearly,
polarized so the changes in the plane of polarization could be eventually of order one. The effect is independent of the frequency.
Equation \eqref{likelyrotat} shows however that the process of rotation due to the CAB is very slow, with a
characteristic time $\eta_0^{-1}$.

Typically in interferometric-type experiments the laser light is made to bounce and folded many times.
Formula \eqref{likelyrotat} can be used each time that the light travels back and forth. When this happens, $\beta$ changes
sign and so does $\sin2\beta$. Since $\epsilon$
is always negative, the effect of the magnetic field is always to increase $\beta$ in absolute value
(i.e. moving the polarization plane away from the magnetic field). So
in this sense, the rotation accumulates.
The situation is different for the CAB term. It does not change sign when $\beta$ does, so its effect
compensates each time the light bounces. However, recall that $\eta_0$
changes sign with a half-period $\pi m_a^{-1}$ so the effect could be accumulated by tuning the length
between each bounce. The range of values of $\pi m_a^{-1}$ makes
this perhaps a realistic possibility for table-top experiments (we are talking here about separations between
the mirrors ranging from millimeters to meters for most accepted values of $m_a$).

It turns out that for Peccei-Quinn axions the effect is actually independent both of the actual values for
$f_a$ and $m_a$ and it depends only on the combination $f_a m_a \simeq 6 \times 10^{15}$ eV$^2$ and the local
axion density. Assuming that the laser beam travels a distance $L=\pi m_a^{-1}$ before bouncing, the total
maximum rotation athat can be observed will be given by $|\eta_0| x$. The total travelled distance will be
$x= {\cal N} L$, where ${\cal N}$ is the total number of turns that depends on the finesse
of the resonant cavity. Replacing the expression for $|\eta_0|$ in the previous expression in terms of the local
DM density $\rho$ (that we assume to be 100\% due to axions) we get
$|\eta_0|= g_{a\gamma\gamma} \frac{4\alpha}{\pi^2}\frac{\sqrt{2\rho}}{f_a}$. Then
\be
|\eta_0| x =  g_{a\gamma\gamma} \frac{4\alpha}{\pi}\frac{\sqrt{2\rho}}{6 \times 10^{15}\ {\rm eV}^2} {\cal N}
\simeq 2 \times 10^{-18} \ {\rm eV}^{-2} \times \sqrt\rho \times {\cal N}.
\ee
Plugging in the expected value for the local axion density one gets for every bounce an increment
in the angle of rotation of $2 \times 10^{-20}$. This is of course a very small number and we realize that the chances
of being able to measure this anytime soon are slim. At present there are cavities whose 
reflection losses are below 1 ppm\cite{supermirrors} but these numbers still fall short.
However this result may be interesting for several reasons.
First of all, it is actually independent from the axion parameters, as long as they are Peccei-Quinn axions, except
for the dependence on $g_{a\gamma\gamma}$ that is certainly model dependent but always close to 1. Second,
in this case it depends directly on the local halo density and nothing else. Third, a positive result obtained by
adjusting the length of the optical path would give an immediate direct measurent of $m_a$ and an indirect
one of $f_a$. There are no hidden or model dependent assumptions, the only ingredient that is needed is QED.

Observing a net rotation of the initial plane of polarization when the magnetic field is absent
(or very small) would be a clear signal of the collective effect of a CAB. On the contrary, a non-zero value
for $\eta_0$ does not contribute at leading order to a change in the ellipticity (and subleading corrections
are very small). In Ref. \cite{ahlers} the authors discuss in some detail the different backgrounds, all of wich are very small
with the exception of the dichroism originating from the experimental apparatus itself \cite{zavattini}. Ways of partially
coping with these experimental limitations are discussed in the previous reference.

Notice that the effect is directly proportional to the distance travelled and
therefore any improvement in the finesse of the cavity directly translates into a longer distance and a better
bound. Recall that in order to measure the rotated angle it is actually much better not to consider an external
magnetic field, making the experimental setup much easier. Incidentally this also liberates us from
the non-linear QED effects discussed in section \ref{sec:eom}.

Axion-like particles not constrained by the Peccei-Quinn relation $f_am_a \simeq$ constant could be easier to rule out
if they happen to be substantially lighter than their PQ counterparts as cavities in this case can be longer and
one could have longer accumulation times.

\section{Conclusions}
In this work we have extended the analysis of axion-photon mixing in the presence of an external magnetic field to the case
where a cold axion background (CAB) is present too. The mixing is then substantially more involved and the two photon
polarizations mix even without a magnetic field. In particular in our results we can take the limit where the magnetic field
vanishes, a situation that would make experiments easier even if it would be really challenging to measure the predicted
effects.
Together with resonant cavity experiments, such as ADMX, optical experiments or observations are so far the only ones
that appear eventually capable
of testing the nature of the CAB.

We have made one approximation that we believe is not essential, namely we have approximated the assumed sinusoidal
variation in time of the CAB by a piece-wise linear function; resulting in a fully analytically solvable problem. We believe
that this captures the basic physics of the problem and we expect only corrections of ${\cal O}(1)$ in
some numerical coefficients
but no dramatic changes in the order-of-magnitude estimates.

The existence of some momentum gaps due to the periodic time dependence of the CAB and its implications has been reviewed too.
It seems challenging to design experiments to verify or falsify their existence, but in any case they are unavoidable if
dark matter is explained in terms of an axion background; in fact it would possibly be the most direct evidence
of the existence of a CAB.

We have obtained the proper modes and their ellipticities and we have analyzed in detail the evolution
of the system. It should be said that CAB-related effects dominate in some regions of the allowed parameter space.
We have also studied the possible presence of accumulative effects that might enhance the rotation
of the instantaneous plane of polarization. This would also be a genuine CAB effect.

In order to analyze the evolution of the system we have made use of the two point function for the
electric field, that correlates the value at $x=0$ with the one at a given value for $x$. We find this a convenient
and compact way of treating this problem. It is valuable to have this tool at hand as the propagator encompasses
all the information of the travelling photons.

Of course the most relevant question is whether laser experiments
may one day shed light on the existence and properties of the CAB. The present authors are not competent to
judge on the future evolution of the precision in this type of experiments. In both cases the required precision is several
orders of magnitude beyond present accuracy, but progress in this field is very fast.

Apart from the precision issue, there are several caveats to take into account when attempting to
experimentally test the predictions of the present work. For instance, a scan on
$m_a$ (i.e. the mirror separation) has to be performed until a cumulative effect is found, which obviously takes time
(this is in a sense somewhat equivalent to the scan on the resonant frequency of the cavity in ADMX).
The total number of reflections
is limited by mirror quality (finesse) and it typically induces a spurious rotation that needs to be disentangled from
the true effect. We do not think that any of the approximations made in this work (basically the piecewise linear
 approximation
for the CAB profile) is experimentally significant provided that the coherence length of the CAB is larger than
the spatial region experimentally probed.

As emphasized in the introduction, checking the coherence of a putative cold axion background is not easy because
the physical effects associated to it are subtle and small in magnitude. The present proposal analyzes the consequences of
the existence of a CAB on photon propagation and as we have seen its effects can be of a size comparable
to other phenomena that are being actively investigated in optical experiments. For these reasons we believe it is important
to bring the present analysis to the
atention of the relevant experimental community.

\section*{Acknowledgements}
This work is supported by grants FPA2010-20807, 2009SGR502 and Consolider grant CSD2007-00042 (CPAN). A. Renau acknowledges the
financial support of a FPU pre-doctoral grant. It is a pleasure to thank several of the participants in the 9th Patras Workshop
for discussions. The authors are grateful to G. Cantatore for discussions on polarimetric experiments.

\appendix
\section{Propagator}
Considering only the spatial components, eq. (52) of Ref. \cite{last} becomes:
  \begin{eqnarray}
  \mathcal{D}^{ij}(\omega,k)&=&D^{ij}+i\omega^2\Bigg\{\frac{b^ib^j}{(k^4-\eta_0^2\vec k^2)(k^2-m_a^2)-\omega^2k^2b^2}\cr
  &&+\frac{i\eta_0k^2(b^iq^j-q^ib^j)}{(k^4-\eta_0^2\vec k^2)[(k^4-\eta_0^2\vec k^2)(k^2-m_a^2)-\omega^2k^2b^2]}\Bigg\},
  \end{eqnarray}
  where
  \begin{equation}
  D^{ij}=-i\left(\frac{P_+^{ij}}{k^2-\eta_0|\vec k|}+\frac{P_-^{ij}}{k^2+\eta_0|\vec k|}\right),\quad \vec q=(\vec b\times\vec k)
  \end{equation}
and the projectors $P_\pm$ have been defined in Ref. \cite{sasha}.
Terms proportional to $k^ik^j$ have been dropped, since we are interested in contracting the propagator with a photon polarization vector.
The roots of the denominators are $|\vec k|=F_j$, with
\begin{eqnarray}
F^2_{1,2}&=&\omega^2+\frac{\eta_0^2}2\mp\frac{\eta_0}2\sqrt{4\omega^2+\eta_0^2}\approx\omega^2\mp\omega\eta_0,\cr
F^2_{3,4}&=&\omega^2-\frac{m_a^2-\eta_0^2}{3}+\sqrt{W}(\cos\chi\mp\sqrt3\sin\chi),\cr
F^2_5&=&\omega^2-\frac{m_a^2-\eta_0^2}{3}-2\sqrt{W}\cos\chi.
\end{eqnarray}
\begin{eqnarray}
W&\approx&\left(\frac{m_a^2}3\right)^2\left(1+\frac{3\omega^2b^2}{m_a^4}\right),\cr
\chi&\approx&\frac{1}{m_a^2}\sqrt3\omega\xi,\cr
\xi&\approx&\left(1+\frac{9\omega^2b^2}{2m^4}\right)^{-1}\sqrt{\eta_0^2+
\left(\frac{\omega b^2}{2m^2}\right)^2+\left(\frac{\omega^2 b^3}{m_a^4}\right)^2}.
\end{eqnarray}
$F_1$ and $F_2$ correspond to the pieces with $P_+$ and $P_-$, respectively.
The piece proportional to $b^ib^j$ has poles at $F^2_{3,4,5}$ and the last piece contains all five poles.
We decompose the denominators in simple fractions:
\begin{equation}
\frac1{(k^4-\eta_0^2\vec k^2)(k^2-m_a^2)-\omega^2k^2b^2}=\sum_{l=3}^5\frac{A_l}{\vec k^2-F_l^2},
\end{equation}
with
\begin{equation}
A_l=\frac{-1}{\prod_{m\neq l,1,2}(F_l^2-F^2_m)},\quad l=3,4,5
\end{equation}
and
\begin{equation}
\frac{k^2}{(k^4-\eta_0^2\vec k^2)[(k^4-\eta_0^2\vec k^2)(k^2-m_a^2)-\omega^2k^2b^2]}=\sum_{l=1}^5\frac{\tilde A_l}{\vec k^2-F_l^2},
\end{equation}
with
\begin{equation}
\tilde A_l=\frac{-(\omega^2-F_l^2)}{\prod_{m\neq l}(F_l^2-F^2_m)},\quad l=1,...,5.
\end{equation}
Then,
\begin{eqnarray}
\mathcal{D}^{ij}(\omega,\vec k)&=&i\left(\frac{P_+^{ij}}{\vec k^2-F^2_1}+\frac{P_-^{ij}}{\vec k^2-F^2_2}\right)\cr
&&+i\omega^2b^2\left[\hat b^i\hat b^j\sum_{l=3}^5\frac{A_l}{\vec k^2-F^2_l}+
i\eta_0(\hat b^i\hat q^j-\hat q^i\hat b^j)\sum_{l=1}^5\frac{|\vec k|\tilde A_l}{\vec k^2-F^2_l}\right]
\end{eqnarray}
We choose the axes so that
\begin{equation}
\hat k=(1,0,0),~\hat b=(0,1,0),~\hat q=(0,0,-1).
\end{equation}
The propagator in position space is, after dropping an overall factor,
\begin{eqnarray}
d^{ij}(\omega,x)&\approx&(P_+^{ij}+P_-^{ij})\cos\left(\frac{\eta_0x}2\right)+i(P_+^{ij}-P_-^{ij})\sin\left(\frac{\eta_0x}2\right)\cr
&&+\hat b^i\hat b^j\sum_{l=3}^5a_le^{i\alpha_lx}-i(\hat b^i\hat q^j-\hat q^i\hat b^j)\sum_{l=1}^5\tilde a_le^{i\alpha_lx},
\end{eqnarray}
where
\begin{equation}
a_l=\frac{\omega^3b^2A_l}{F_l},\quad\tilde a_l=\omega^3b^2\eta_0\tilde A_l,\quad\alpha_l=F_l-\omega.
\end{equation}
All the $\alpha_l$ are proportional to $\eta_0$ or $b^2$, except for $\alpha_5\approx-\frac{m_a^2}{2\omega}$.
Restricting ourselves only to $y-z$ components, we can write $d(\omega,x)$ in matrix form.
\begin{equation}
P^i_{+j}+P^i_{-j}=
\left(
  \begin{array}{cc}
    1 & 0 \\
    0 & 1 \\
  \end{array}
\right),
\end{equation}
\begin{equation}
i(P^i_{+j}-P^i_{-j})=
\left(
  \begin{array}{cc}
    0 & 1 \\
    -1 & 0 \\
  \end{array}
\right),
\end{equation}
\begin{equation}
\hat b^i\hat b_j=
\left(
  \begin{array}{cc}
    -1 & 0 \\
    0 & 0 \\
  \end{array}
\right),
\end{equation}
\begin{equation}
-i(\hat b^i\hat q_j-\hat q^i\hat b_j)=
\left(
  \begin{array}{cc}
    0 & -i \\
    i & 0 \\
  \end{array}
\right).
\end{equation}
If we write
\begin{equation}
\sum_l a_le^{i\alpha_lx}=-(\epsilon+i\varphi),\quad i\sum_l \tilde a_le^{i\alpha_lx}=-(\tilde\epsilon+i\tilde\varphi),
\end{equation}
we have
\begin{equation}\label{propmatrix}
d^i_j(\omega,x)=
\left(
  \begin{array}{cc}
    \cos\frac{\eta_0x}{2}+\epsilon+i\varphi & \sin\frac{\eta_0x}{2}+\tilde\epsilon+i\tilde\varphi \\
    -\left(\sin\frac{\eta_0x}{2}+\tilde\epsilon+i\tilde\varphi\right) & \cos\frac{\eta_0x}{2}\\
  \end{array}
\right)
\end{equation}

\section{Ellipticity and rotation}
The quantities appearing in \eqref{propmatrix} are
\be
\epsilon\approx-\frac{\omega^2b^2}{m_a^4}\left(1-\cos\frac{m_a^2x}{2\omega}\right),\quad
\varphi\approx\frac{\omega^2b^2}{m_a^4}\left(\frac{m_a^2x}{2\omega}-\sin\frac{m_a^2x}{2\omega}\right),
\ee
while $\tilde\epsilon$ and $\tilde\varphi$ are both proportional to $b^2\eta_0$, so they are negligible.\\
In the limit $\frac{m_a^2x}{2\omega}\ll1$ we have
\begin{equation}\label{lim1}
\epsilon\approx -\frac{b^2x^2}8,\quad\varphi\approx \frac{m_a^2b^2x^3}{48\omega}
\end{equation}
whereas if $\frac{m_a^2x}{2\omega}\gg1$ the trigonometric functions oscillate rapidly and can be dropped:
\begin{equation}\label{lim2}
\epsilon\approx-\frac{\omega^2b^2}{m_a^4},\quad\varphi\approx\frac{\omega b^2x}{2m_a^2}.
\end{equation}
Eq. (\ref{lim1}) agrees with eq. 16 of Ref. \cite{mpz}  (although their $k^2$ in the denominator should be only $k$,
the dimensions do not fit otherwise).
Eq. (\ref{lim2}) agrees with their eq. (20,21), at least to second order in $b$.

If we start with a polarization $\vec n_0=(\cos\beta,\sin\beta)$, after a distance $x$ we have
\begin{equation}
n^i_x=d^i_j(x)n^j_0=\left(\begin{array}{c}
\cos(\beta-\frac{\eta_0x}{2})+(\epsilon+i\varphi)\cos\beta\\
\sin(\beta-\frac{\eta_0x}{2})
                                  \end{array}\right)
\end{equation}
Following section 1.4 of Ref. \cite{bornwolf}, this vector describes a polarization at an angle
\begin{equation}
\alpha\approx\beta-\frac{\eta_0x}{2}-\frac\epsilon2\sin2\beta
\end{equation}
and with ellipticity
\begin{equation}
e=\frac12\left|\varphi\sin2\beta\right|.
\end{equation}
This ellipticity differs from the one described in Ref. \cite{mpz} by the factor of $\sin2\beta$.

Quantum mechanically the quantity that is relevant is not the amplitude itself, but the modulus
squared of it. From this, the
probability of finding an angle $\alpha$ given an initial angle $\beta$ will be
\begin{equation}
P(\alpha,\beta)=\left|\epsilon'_id^{ij}\epsilon_j\right|^2\approx\cos^2\left(\alpha-\beta+\frac{\eta_0x}{2}\right)
+2\epsilon\cos\left(\alpha-\beta+\frac{\eta_0x}{2}\right)\cos\alpha\cos\beta.
\end{equation}
The angle of maximum probability, satisfying $\partial_\alpha P(\alpha,\beta)=0$ is also, to first order,
\be
\alpha=\beta-\frac{\eta_0x}{2}-\frac\epsilon2\sin2\beta.
\ee

%\begin{thebibliography}{000} %for 3 digits
%\begin{thebibliography}{00}  %for 2 digits

\end{document}